\documentclass{PoS}

\usepackage{wrapfig}
\usepackage{lipsum}
\usepackage{float}
\usepackage{subcaption}
\title{Dark matter gamma-ray line searches toward the Galactic Center halo with H.E.S.S. I}

\ShortTitle{DM $\gamma$-ray lines search at the GC}

\author{\speaker{Lucia Rinchiuso$^1$, Emmanuel Moulin$^1$}, Aion Viana$^2$, Christopher Van Eldik$^3$, Johannes Veh$^3$ for the H.E.S.S. collaboration\\%\thanks{A footnote may follow.}\\
        $^1$ IRFU, CEA, Universit\'e Paris-Saclay, F-91191 Gif-sur-Yvette, France\\
        E-mail: \email{lucia.rinchiuso@cea.fr}, \email{emmanuel.moulin@cea.fr}\\
        $^2$ Particle Physics and High-Energy Astrophysics Division, MPIK, 69029 Heidelberg, Germany\\
        $^3$ Erlangen Centre for Astroparticle Physics, Erwin-Rommel-Str. 1, 91058 Erlangen, Germany}

\abstract{The presence of dark matter is nowadays widely supported by a large body of astronomical and cosmological observations. A large amount of dark matter is expected to be present in the central region of the Milky Way. Very-high-energy (>100 GeV) $\gamma$-rays can be produced in the annihilation of dark matter particles. The H.E.S.S. array of Imaging Atmospheric Cherenkov Telescopes is a powerful tools to observe the Galactic Centre trying to detect $\gamma$-rays from dark matter annihilation. A new search for a dark matter signal has been carried out on the full H.E.S.S.-I data set of 2004-2014 observations. A 2D-binned likelihood method has been applied to exploit the spectral and spatial properties of signal and background. Updated constraints are derived on the velocity-weighted annihilation cross section for signals from prompt annihilation of dark matter particles into two photons. The larger statistics from the 10-year Galactic Center dataset of H.E.S.S.-I together with the 2D-analysis technique allows to significantly improve the previous limits.}

\FullConference{35th International Cosmic Ray Conference – ICRC217-\\
		10-20 July, 2017\\
		Bexco, Busan, Korea}

\begin{document}

\section{Dark Matter toward the Galactic Center}
There is today a substantial body of evidences for the existence of Dark Matter (DM). Its gravitational evidence has been suggested in the early '30s to explain velocities of galaxies in clusters and in the '70s  from the measurements of the  rotation curves of galaxies.  In the era of precision cosmology measurements, it is now possible to estimate precisely the dark matter density. These measurements show that about the $25\%$ of the total content of the Universe is made out of cold DM in the standard model of cosmology known as the  $\Lambda$CDM model \cite{bib:planck}. Among the most promising particle candidates are the weakly interacting massive particles (WIMPs): massive particles with mass scale and coupling strengths at the electroweak scale \cite{bib:WIMP}.   
Interestingly, if thermally-produced in the early Universe, such particles have naturally a self-annihilation cross section that is expected from cosmological measurements.
\subsection{Dark matter indirect search} \label{sub:DM}
DM can be indirectly detected through the measurement of its self-annihilation products. Among them are  very-high-energy (VHE, E$>$100 GeV) $\gamma$-rays that can be detected with ground-based  arrays of Cherenkov telescopes like H.E.S.S. \cite{bib:HESS}. Since $\gamma$-rays are not bent in Galactic magnetic fields and thus point back to their production site, these experiments would be able to recover the distribution of DM through the reconstruction of  direction of $\gamma$-rays and estimate the DM mass by reconstructing their energy spectrum. 
The strategy for DM indirect detection is to look in regions in the sky with strong annihilation signal, {\it i.e.} proximity and large amount of DM, and low contamination from other VHE $\gamma$-rays astrophysical sources. For these reasons, the Inner Galactic halo and the dwarf galaxy satellites of the Milky Way are considered as the most promising targets for indirect detection with $\gamma$-rays. Indeed, the former is expected to host the largest amount of DM, despite the presence of important astrophysical background, while the latters are the most DM-dominated objects in the universe and free from VHE background so far. However, the DM signals from dwarf galaxies are expected to be much fainter than from the Galactic Centre (GC) region, and thus more difficult to detect. The GC region is the best target for DM detection with VHE $\gamma$-rays.

The flux of $\gamma$-rays from DM annihilation writes as 
\begin{equation}
\frac{d\phi(\Delta\Omega,E)}{dE}=\frac{1}{4\pi}\frac{\langle\sigma v\rangle}{2m_{\rm DM}^2}\sum Br_{\rm i}\frac{dN_{\rm i}(E)}{dE}\times J(\Delta\Omega)\, .
\end{equation}
$m_{\rm DM}$ and $\langle\sigma v\rangle$ are the mass and the thermally-averaged velocity-weighted annihilation cross section of the DM particle, respectively.  $\frac{dN_{\rm i}}{dE}$ is the $\gamma$-ray spectrum for the annihilation channel  $i$  with the  branching ratio $Br_{\rm i}$. The distribution of the DM density is contained in the J-factor $J(\Delta\Omega)$, which is the integral of the DM density squared along the line of sight (los) and over the solid angle $\Delta\Omega$: 
\begin{equation}
\label{eq:jfactor}
J(\Delta\Omega)=\int_{\Delta\Omega}\int_{\rm los}\rho^2(s,\theta)dsd\Omega.
\end{equation}
$s$ is the coordinate along the line of sight and $\theta$ the angle between the direction of observation and the Galactic plane. 
\subsection{Dark matter spectrum and density profile} \label{sub:ij}
The self-annihilation of DM can give rise to two peculiar energy spectra, referred hereafter as to the  continuum and $\gamma$-ray line spectra, respectively. The continuum of VHE $\gamma$-rays \cite{bib:continuum} results from the annihilation into W/Z bosons, quarks or leptons and their subsequent hadronization and/or decay. The $\gamma$-ray line comes from the prompt annihilation of DM particles into two photons. This process is suppressed compared to the continuum one because it cannot take place at tree-level. However, it gives the clearest signature of presence of DM. 
\begin{figure*}[htbp]
    \centering
    \hspace{-0.5cm}
    \begin{subfigure}[t]{0.6\textwidth}
        \centering
        \includegraphics[height=6.5cm]{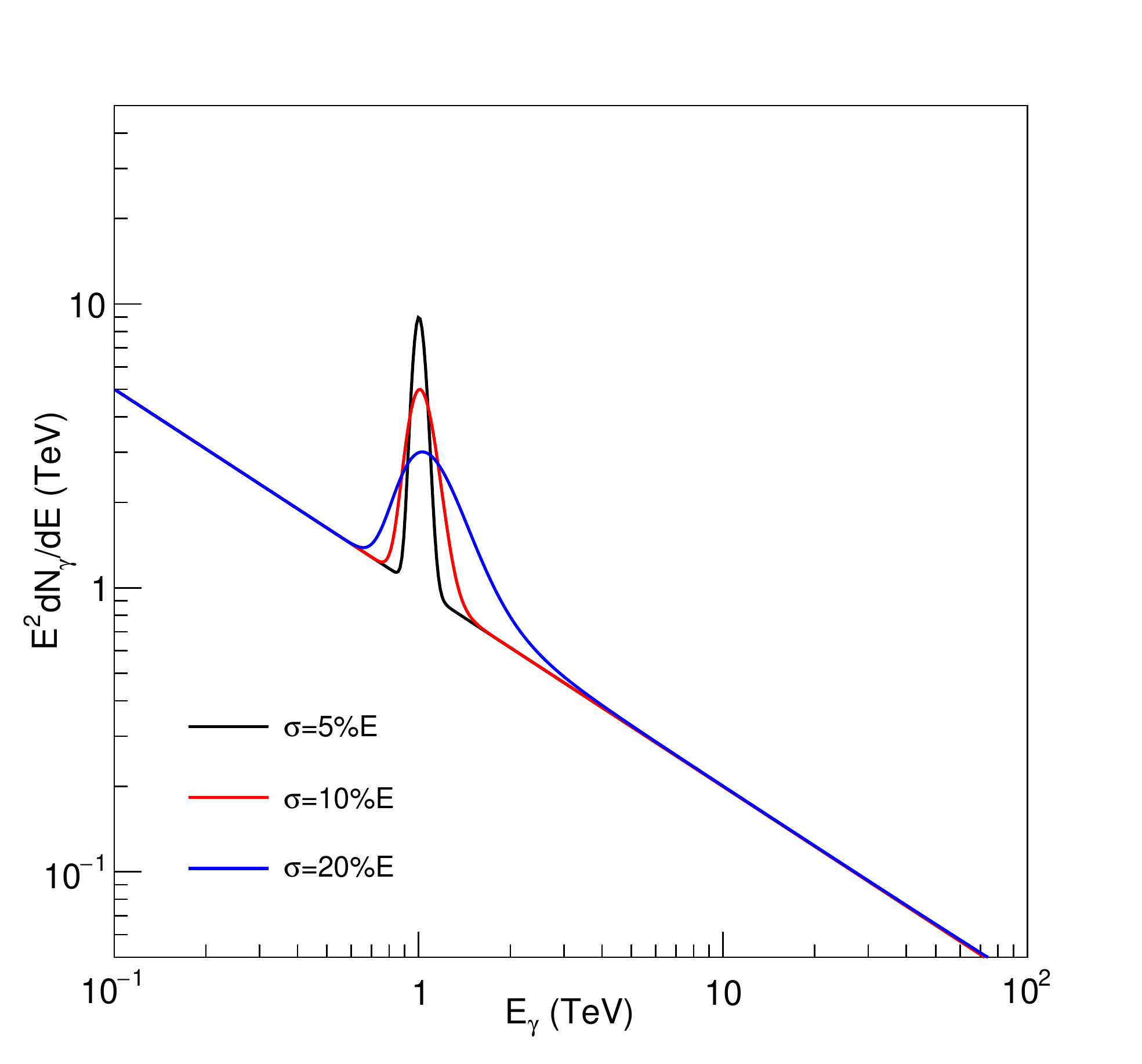}
        \caption{}
	\label{fig:spectrum}
    \end{subfigure}%
    ~\hspace{-2.5cm}
    \begin{subfigure}[t]{0.6\textwidth}
        \centering
        \includegraphics[height=6.5cm]{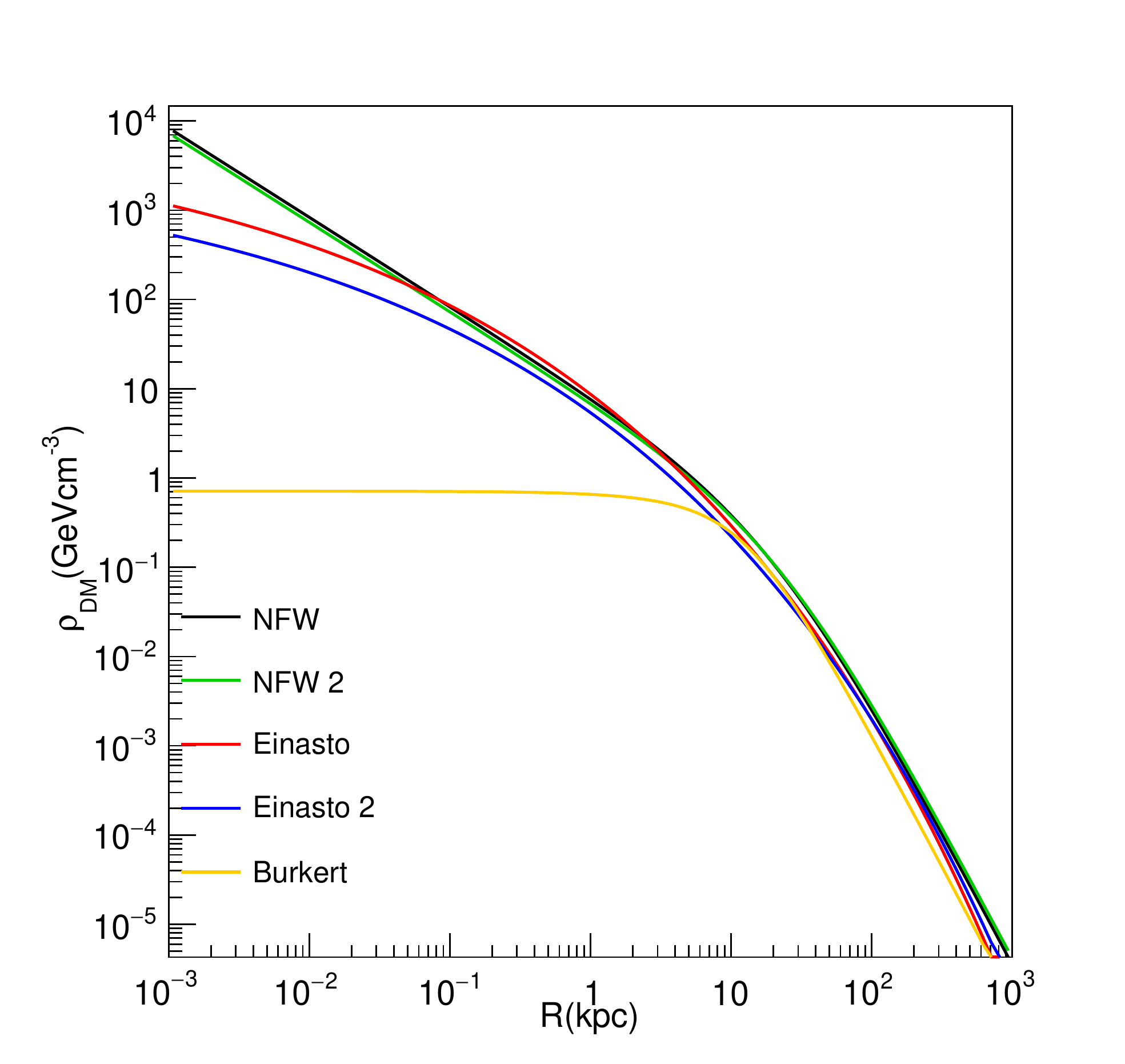}
        \caption{}
	\label{fig:DMprofile}
    \end{subfigure}
\caption{{\it Left:}  line spectrum from prompt annihilation of WIMPs into two $\gamma$-rays convluted by Gaussian energy resolution of $5\%$ (black solid line), $10\%$ (red solid line) and $20\%$ (blue solid line) of energy. {\it Right:} Examples of DM density profiles in the Milky Way as a function of the distance from the GC. The Einasto profile (solid red line) has been chose for this analysis.}
\end{figure*}
$\gamma$-ray line spectra are shown in Fig. \ref{fig:spectrum} superimposed to a power-law background of index 2.7. The DM line is spread with a $5\%$ (black solid line), $10\%$ (red solid line) and $20\%$ (blue solid line) energy resolution to show that a good resolution is crucial for a better signal detection. The thermal relic cross section, that is aimed to be probed with experiments, is $\sim3\times10^{-26}$ cm$^3$s$^{-1}$ for the continuum and of the order of $10^{-29}$ cm$^3$s$^{-1}$ for the $\gamma$-ray line. The distinctive spectral feature of the DM annihilation signal with respect to the residual background can be exploited to improve signal-to-background discrimination.\\
The strength of DM signal also depends strongly on the DM density profile. The region below 1~kpc around the GC is expected to host a large amount of DM. However, its spatial distribution is not well constrained. Hence, several DM density profiles are assumed in the Inner Galactic halo. Among them two kinds of parametrizations are used: cored and cuspy profiles. Fig. \ref{fig:DMprofile} shows some of these profiles as a function of distance $R$ from the GC. 
The coordinate $R$ is related to $s$ as $R = \sqrt{s^2 - 2r_{\odot}s \,cos \theta + r_{\odot}^2}$, where $r_{\odot}$  is the distance of the observer to the GC assumed to be 8.5~kpc.
The cored profiles - {\it e.g.} Burkert profile \cite{bib:Burkert}, in Fig. \ref{fig:DMprofile} (yellow solid line) - become flat towards the GC, while the cuspy profiles steaply increase towards the Inner Galactic halo. Examples of cuspy profiles are the Einasto \cite{bib:Einasto} and Navarro, Frenk and White (NFW) \cite{bib:NFW} profiles, shown with different parametrizations in Fig. \ref{fig:DMprofile}. For this analysis the Einasto profile expressed as  
\begin{equation}
\rho_{\rm Ein}(R)=\rho_s\exp{\bigg[\frac{-2}{\alpha}\bigg(\Big(\frac{R}{r_s}\Big)^{\alpha}-1\bigg)\bigg]}
\end{equation}
is used, with a standard parametrization as in the previous search for DM $\gamma$-ray lines toward the GC in 2013 \cite{bib:2013}. 
This profile shows a strong gradient of DM  towards the GC and it is well suited to take advantage of the spatial discrimination between the signal and the residual background distributions, the latter being isotropically distributed.

\section{Data analysis with H.E.S.S.-I observations towards the Galactic Center region}
\subsection{The 10-year observational data set and the regions of interest for DM searches}
This analysis has been performed on the data set collected by the four 12 m diameter telescopes of H.E.S.S. \cite{bib:HESS} from 2004 to 2014, during the first phase of the experiment. A total of 254 hours of observations were available at the nominal position of the GC. The dataset has an averaged zenith angle of $19^\circ$ and is obtained from observational pointings between $0.5^\circ$ and $1.5^\circ$ in radial distance from the GC.

\begin{figure}[htbp] 
  \begin{center}
    \includegraphics[width=0.8\textwidth]{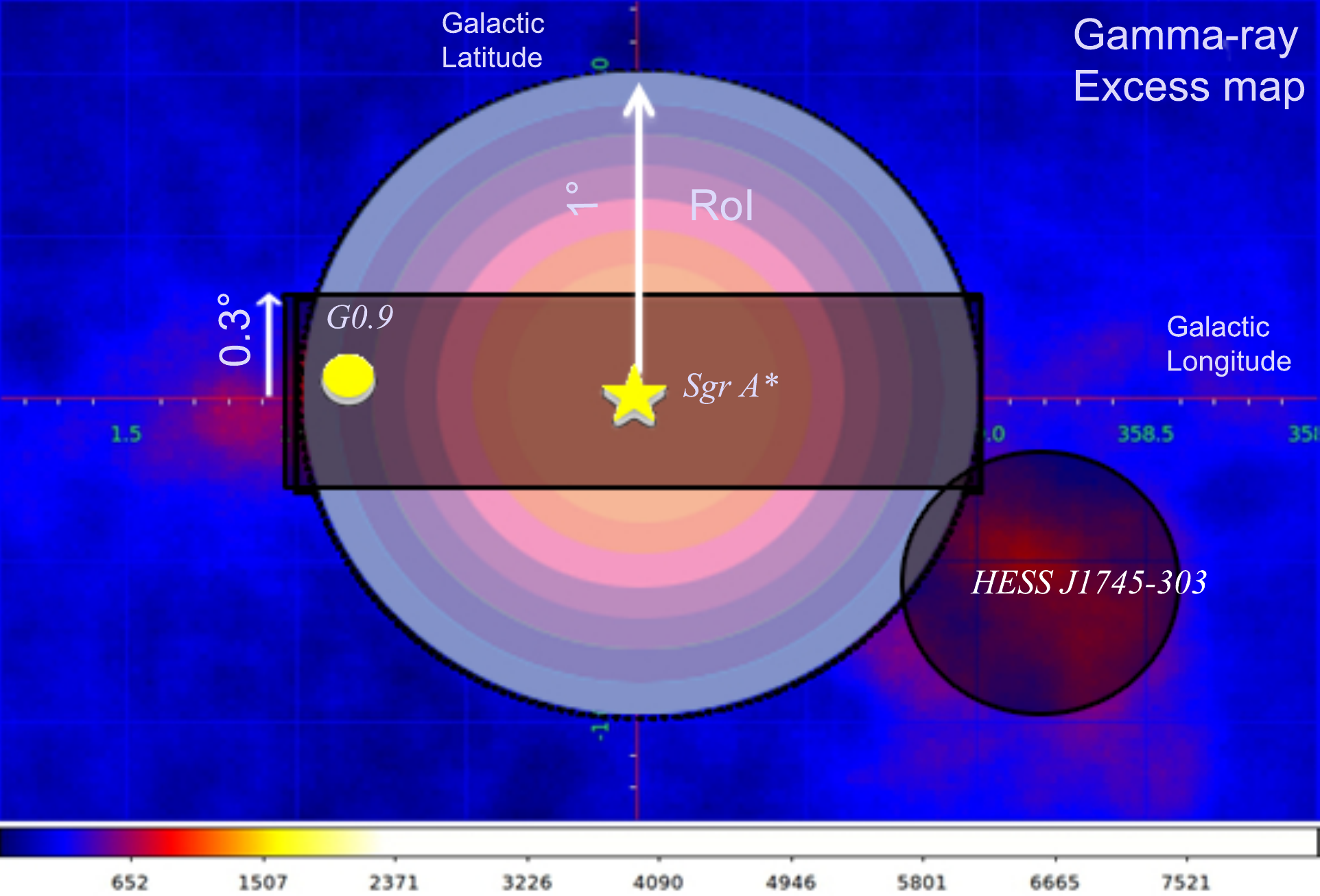}
    \caption{VHE $\gamma$-ray excess map in Galactic coordinates of the inner 300 pc of the Milky Way seen by H.E.S.S. Bright sources and diffuse emission are observed. The ON region for DM search is defined as a circle of $1^\circ$ centered at the GC and it is divided in 7 sub-regions of interest. The band of $\pm0.3^{\circ}$ around the Galactic Plane is excluded for the data analysis. A circle of radius $0.4^\circ$ centered on HESS J1745-303 is excluded as well.}
    \label{fig:GC}
  \end{center}
\end{figure} 
The VHE $\gamma$-ray excess map of the observed region is shown in Fig. \ref{fig:GC}. Over the map the region of interest (RoI) where the signal is looked for, referred to as the ON region, is represented as a circle of radius $1^\circ$ around the GC. This RoI is divided in 7 sub-regions of interest to take advantage of the DM cuspy spatial distribution with respect the isotropic one of the background. The black patches on the map, a box of $\pm 0.3^\circ$ around the galactic plane and a circle of radius $0.4^\circ$ centered in (-$1.29^\circ$,-$0.64^\circ$), are the regions excluded from the data analysis to avoid contamination  in the ON and OFF regions from  standard VHE astrophysical emissions. Indeed, the Inner Galactic halo contains several VHE sources. Among them are HESS J1745-290 coincident with the supermassive black hole Sgr A$^*$ \cite{bib:source290}, HESS J1747-281 coincident with  the pulsar wind nebula G09+0.1 \cite{bib:sourceG09},  the supernova remnant HESS J1745-303 \cite{bib:source303} coincident with G359.1-0.5, and diffuse emission \cite{bib:source290}.
\subsection{Residual background measurement}
The residual background is measured in the OFF regions. For each sub-RoI and each observation the OFF region is build symmetrically to the ON region with respect to the pointing position of the observation. 
The procedure is repeated for the 7 sub-RoIs and all the observation runs. In this way the ON and OFF regions have the same acceptance, due to the azimuthal symmetry of the experiment. Moreover, with a careful symmetric rejection of the excluded regions, the ON and OFF regions end up with the same shape and solid angle size.
In addition, cuspy DM density profiles like the Einasto and NFW profiles allows one to choose ON and OFF regions that both fall in the field of view of the camera ($\sim5^\circ$ in diameter), so that $\gamma$-ray measurements in the ON and in the OFF regions  take place in the same observational and detector conditions. For these DM profiles, one expects a significant DM gradient between the ON and OFF regions. 
%The presence of a larger amount of DM in the ON region with respect to the OFF region is crucial for the analysis. 
Areas of the ON and OFF regions that overlap or where the latter has a higher density of DM than the former, are removed, keeping the same solid angle size. 
The J-factor as written in section \ref{sub:DM} is computed in the same regions as defined for the ON and OFF event measurements. The total J-factor values are obtained for each observation and are weighed by its live time.
\begin{figure}[htbp]
  \begin{center}
    \includegraphics[width=0.65\textwidth]{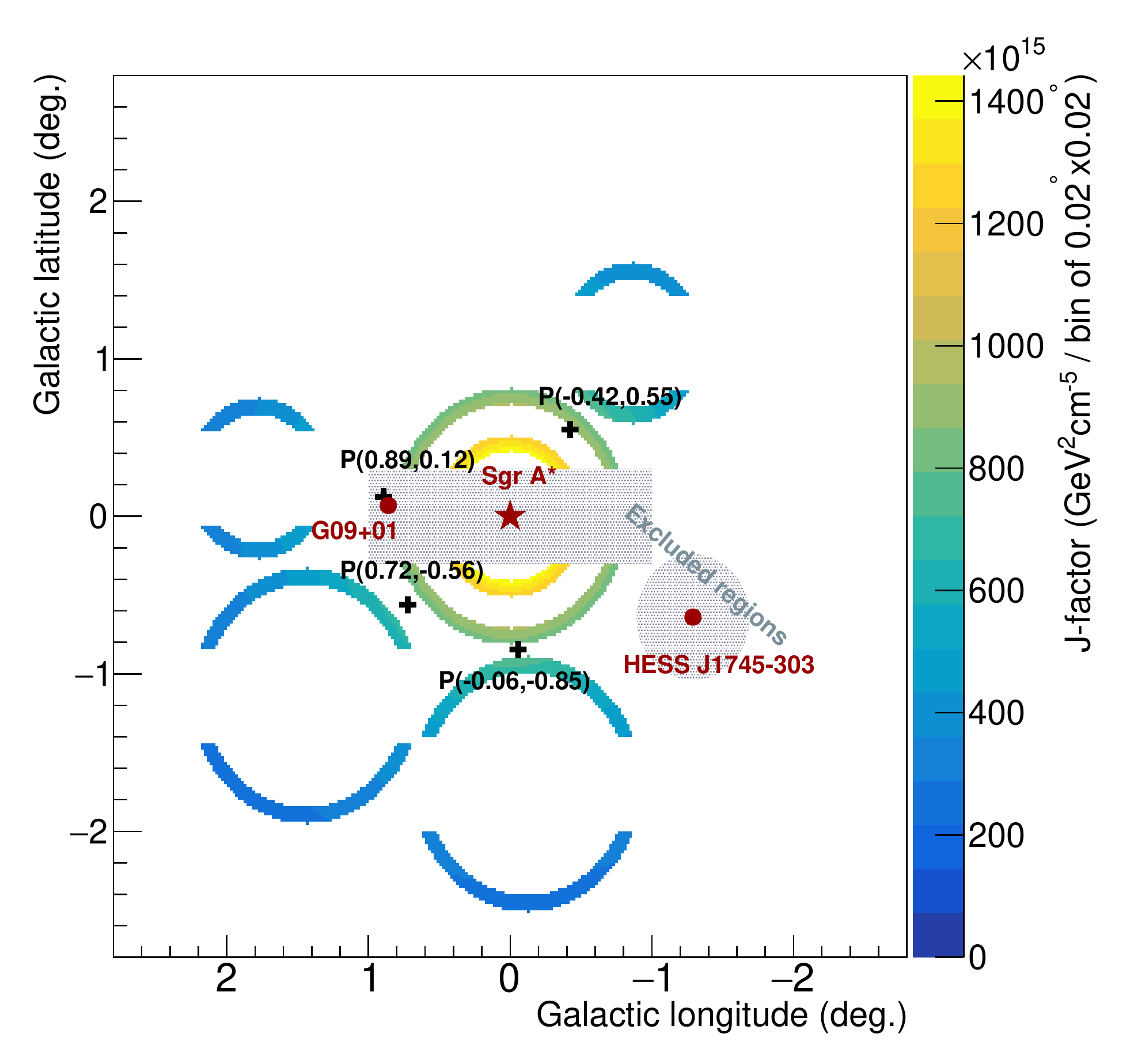}
    \caption{J-factors values in the ON and OFF regions in Galactic coordinates for two different pointing positions (black crosses), and for the 2$^{\rm nd}$ (yellow) and 5$^{\rm th}$ (green) RoIs. The J-factor values are computed for the Einasto profile per bin of $0.02^\circ\times0.02^\circ$. A strong gradient of DM density is observed from the ON to the OFF regions. The excluded regions are shown as grey shaded areas. The main VHE $\gamma$-ray sources are shown : the GC (red star), G0.9 and the SNR HESS J1745-303 (red points).}
    \label{fig:ONOFF}
  \end{center}
\end{figure} 
Fig. \ref{fig:ONOFF} shows the J-factor values for two RoIs in the ON (yellow and green) and OFF (blue) regions for four different pointing positions  in Galactic coordinates. Here, as an example, the sub-RoIs 2 and 5 are shown for two different pointing positions (black crosses). The sources are drawn in red and the grey-shaded patches correspond to the excluded regions. The gradient of DM density between the ON and OFF regions is noticeable. For RoI 2 with the pointing position P(0.89,0.12), a J-factor about a factor of 3 higher is obtained  in the ON compared to the OFF regions.
\subsection{Likelihood data analysis method}
The data analysis method is based on a likelihood ratio test statistic. As performed in the analysis for the continuum annihilation channels in 2016 \cite{bib:continuum}, the likelihood function is binned in 2 dimensions: energy (bins j) and space (bins i, corresponding to the sub-RoIs). The 2D-binned likelihood function is expressed as the product of the Poisson terms for the ON and OFF regions by:
\begin{equation}
\mathcal{L}_{\rm ij}(\mathbf{N_{\rm ON}},\mathbf{N_{\rm OFF},\mathbf{\alpha}}|\mathbf{N_{\rm S}},\mathbf{N_{\rm S}'},\mathbf{N_{\rm B}})=\frac{(N_{\rm S}+N_{\rm B})^{N_{\rm ON}}}{N_{\rm ON}!}e^{-(N_{\rm S}+N_{\rm B})}\frac{(N_{\rm S}'+\alpha N_{\rm B})^{N_{\rm OFF}}}{N_{\rm OFF}!}e^{-(N_{\rm S}'+\alpha N_{\rm B})}.
\end{equation}
$N_{\rm ON}$ and $N_{\rm OFF}$ are the numbers of photons measured in the ON and OFF regions, respectively. The factor $\alpha$ is the ratio between the solid angle size of the OFF and ON regions. $\alpha$=1 by construction. $N_B$ is the background expected in the ON region and it is computed from the requirement $d\mathcal{L}_{\rm ij}/dN_{\rm B}\equiv0$. The number of photons from DM annihilations expected in the ON and OFF regions respectively are $N_{\rm S}$ and $N_{\rm S}'$. The number of photons expected from the signal in the bin (i,j) is computed as 
\begin{equation}
N_{\gamma}(E)=\frac{1}{4\pi}\frac{\langle\sigma v\rangle}{2m_{\rm DM}^2}\int\frac{dN}{dE'}(E')R(E,E')T_{\rm obs}A_{\rm eff}(E')dE'\times J(\Delta\Omega)\, ,
\end{equation} 
where $A_{\rm eff}$ is the energy-dependent effective area and  $T_{\rm obs}$ the observation time. The J-factor is computed for the Einasto profile according to Eq.~(\ref{eq:jfactor}) and the $\gamma$-ray line spectrum is described as a delta function centered in $m_{\rm DM}$: $\frac{dN}{dE'}(E')=2\delta(E'-m_{\rm DM})$. The line is spread with a Gaussian function to account for the finite energy resolution of the experiment with a variance $\sigma/E =10\%$. It expresses $R(E,E')=\frac{1}{\sqrt{2\pi}\sigma}e^{-\frac{(E-E')^2}{2\sigma^2}}$.

The total likelihood in presence of DM, $\mathcal{L}_{\rm w}$, is computed as the product of the $\mathcal{L}_{\rm ij}$ and it is used to construct the test statistics $TS$ defined as $TS=-2\log(\mathcal{L}_{\rm w}/\mathcal{L}_{\rm wo})$,
where $\mathcal{L_{\rm wo}}$ is the likelihood in the null hypothesis, {\it i.e.} without DM : $N_{\rm S}=0$ and $N_{\rm S}'=0$. If no significant excess is observed in the ON region with respect to the OFF region, the value of $TS$ can be used to derive upper limits on $\langle\sigma v\rangle$ at a $95\%$ confidence level (C.L.) by imposing $TS=2.71$ to derive the corresponding value of $\langle\sigma v\rangle$ for a given DM mass.
\section{Results}
No significant $\gamma$-ray excess is observed in any of the 7 RoIs. New limits are derived on the velocity-weighted annihilation cross section for the ${\rm DM DM}\rightarrow\gamma\gamma$ channel, using the full 10-year data set of H.E.S.S.-I observations towards the GC region, the Einasto profile parametrization and the optimized 2D-binned likelihood analysis method.
\begin{figure*}[htbp]
    \centering
    \begin{subfigure}[t]{0.5\textwidth}
        \centering
        \includegraphics[height=7.cm]{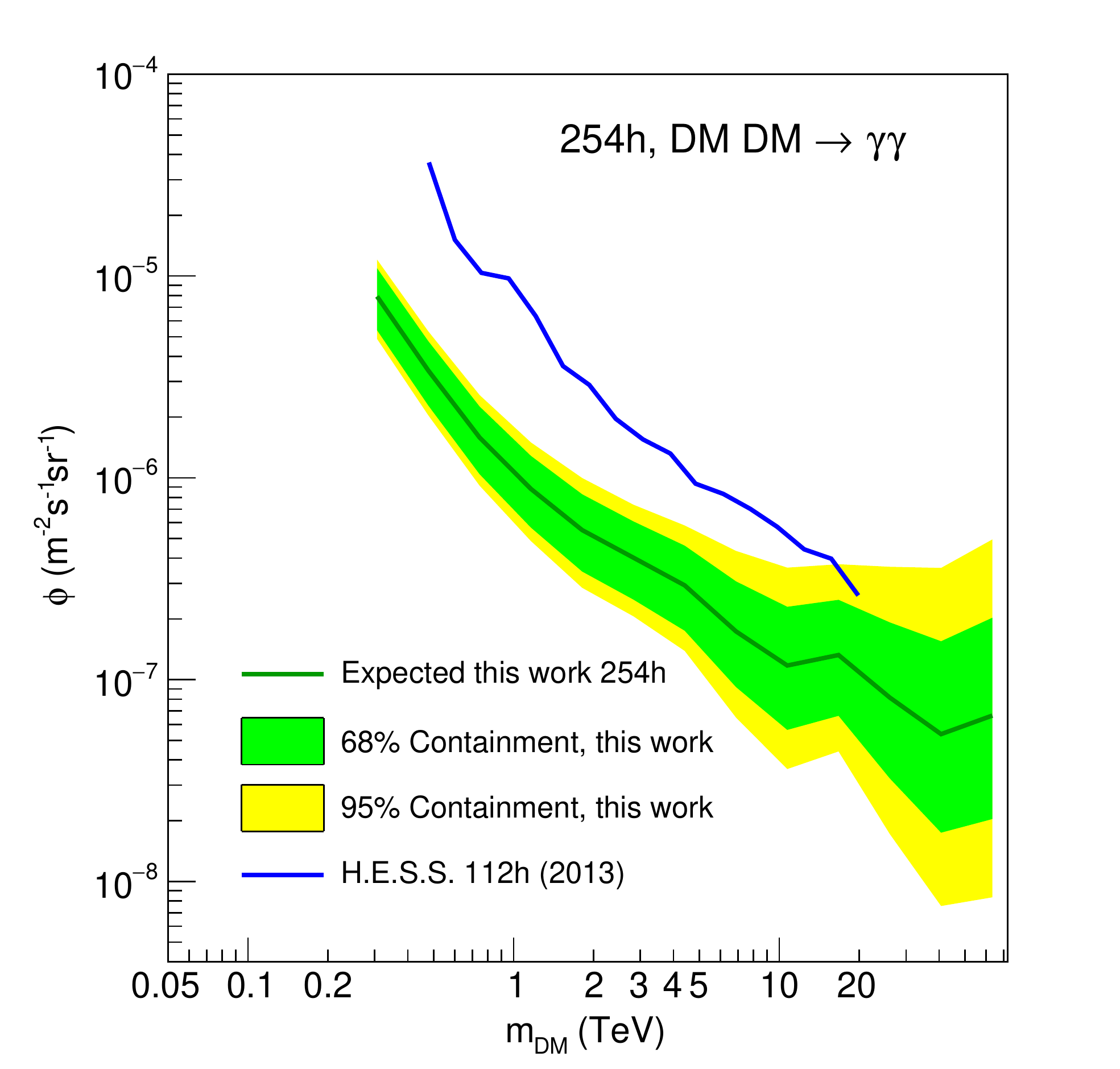}
        \caption{}
	\label{fig:flux}
    \end{subfigure}%
    \hspace{-1cm}~ 
    \begin{subfigure}[t]{0.5\textwidth}
        \centering
        \includegraphics[height=7.cm]{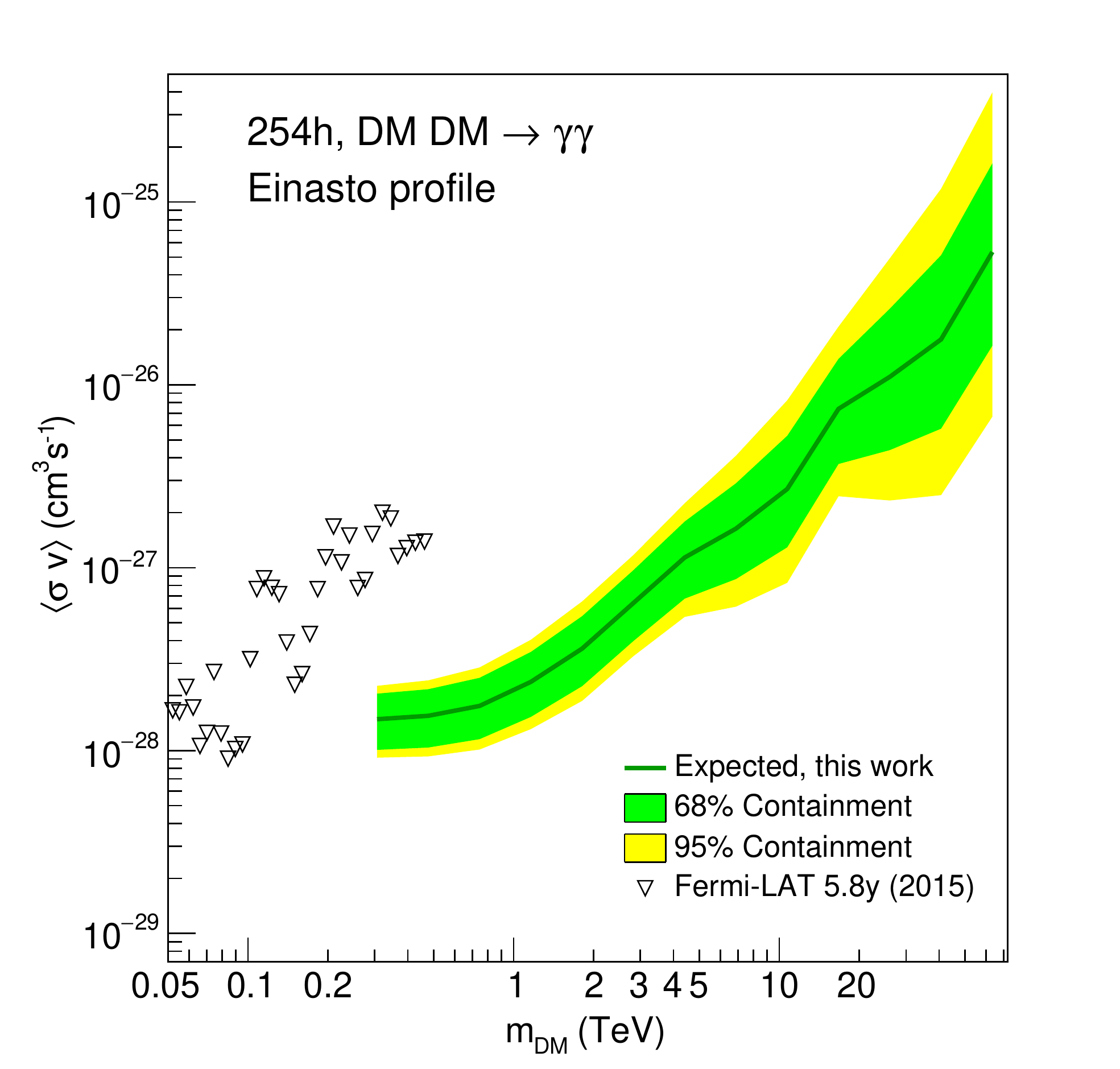}
       \caption{}
	\label{fig:sigma}
    \end{subfigure}
\caption{New constraints on the dark matter annihilation into two photons (${\rm DM DM} \rightarrow\gamma\gamma$)  as a function of the DM mass $m_{\rm DM}$. The  mean expected limits (green solid line) are shown with their 68\% (light green box) and 95\% (yellow box) containment bands.  Previous H.E.S.S. limits published 2013 (blue solid line) are also plotted~\cite{bib:2013}. {\it Left:} 95\% C.L. upper limits on the photons flux $\rm \phi$ from dark matter annihilations. {\it Right:} 95\% C.L. upper limits on the thermally-averaged velocity-weighted annihilation cross section $\langle\sigma v\rangle$ for the Einasto profile, compared to Fermi-LAT limits.}
\end{figure*}
The expected limits on the photon flux  $\rm \phi$ and $\langle\sigma v\rangle$ are shown in Fig. \ref{fig:flux} and Fig. \ref{fig:sigma}, respectively. The mean expected limits are displayed  (green solid line) with their $68\%$ (green area) and $95\%$ (yellow area) containment bands. The behavior like $m_{\rm DM}^2$ is visible at high masses. 
The mean expected limits and their statistical bands are computed using blank field observations. Indeed, the distribution of events in the OFF region is estimated from extragalactic observations, at Galactic latitudes above $10^\circ$, in the same conditions of the observations towards the GC region. 1000 Poisson realisations are then performed from the expected OFF distributions. The likelihood analysis procedure is applied on each realisation in order to build a distribution of the $\langle\sigma v\rangle$ values at $TS=2.71$ for each $m_{\rm DM}$. The mean expected limit is derived from the mean of the distribution and the containment bands from the standard deviation. A systematics uncertainty of $25\%$ is added to account for the uncertainty on the energy resolution. It dominates below $\sim$1~TeV. The best expected limit reaches $1.5\times10^{-28}$ cm$^3$s$^{-1}$ for DM mass of 300 GeV. At 1 TeV, there is an improvement of a factor about 8 with respect to the previous results from 2013 \cite{bib:2013} (blue solid line). 
The improvement  comes from the larger photon statistics, which is doubled with respect to the previously available data set, and the 2D-binned likelihood approach that exploits the spatial information to increase the signal-to-background ratio, and the improved events reconstruction. The events reconstruction performance is comparable to the previous one at high energies, while it is more performant below $\sim1$ TeV.

\section{Summary and outlook} 
The full 10-year data set obtained by H.E.S.S.-I  observations towards the GC has been analyzed looking for DM signatures. No significant excess was observed in the region of interest. New limits have been derived on the thermally-averaged velocity-weighted annihilation cross section of DM particles annihilating into two photons. The expected limits are the most stringent so far in the TeV range. The spectral and spatial DM properties have been exploited for a better discrimination of the searched signal with respect to the residual background. Indeed, the analysis is based on a likelihood function binned in 2 dimensions: energy and space. $95\%$ C.L. mean expected limits have been derived on $\langle\sigma v\rangle$ as a function of the DM particle mass via a likelihood ratio test statistic. The best expected limit reaches $1.5\times10^{-28}$cm$^3$s$^{-1}$ at 300 GeV. This analysis improves over the previous results from 2013 by a factor about 8. The updated H.E.S.S.-I limits also surpass those by Fermi-LAT of a factor about 4 above 300 GeV. This new expected limits are of particular interest for future VHE $\gamma$-ray searches~\cite{bib:enhance} in specific DM particle models with enhanced line signal through the Sommerfeld effect like the Wino dark matter~\cite{bib:wino} and minimal dark matter multiplets~\cite{bib:mdm}.


\begin{thebibliography}{99}
\bibitem{bib:planck} P. A. R. Ade {\it et al.} (Planck Collaboration), A$\&$A 594, A13 (2016).
\bibitem{bib:WIMP} G. Bertone, D. Hopper and J. Silk, Phys. Rep. 405, 793 (2005).
\bibitem{bib:HESS} F. Aharonian {\it et al.} (H.E.S.S. Collaboration), Astron. Astrophys. 457, 899 (2006).
\bibitem{bib:continuum} H. Abdallah {\it et al.} (H.E.S.S. Collaboration), Phys. Rev. Lett. 117, 111301 (2016).
\bibitem{bib:Burkert} P. Salucci and A. Burkert, Astrophys. J. 537:L9-L12 (2000).
\bibitem{bib:Einasto} J. Einasto, Trudy Inst. Astroz. Alma-Ata, No. 17, 1 (1965).
\bibitem{bib:NFW} J. F. Navarro, C. S. Frenk and S. D. M. White, Astrophys. J. 490, 122 (1997).
\bibitem{bib:2013} A. Abramowski {\it et al.} (H.E.S.S. Collaboration), Phys. Rev. Lett. 110, 041301 (2013).
\bibitem{bib:source290} A. Abramowski {\it et al.} (H.E.S.S. Collaboration), Nature 532, 476 (2016).
\bibitem{bib:sourceG09} F. Aharonian {\it et al.} (H.E.S.S. Collaboration), A$\&$A 432, L25 (2005).
\bibitem{bib:source303} F. Aharonian {\it et al.} (H.E.S.S. Collaboration), A$\&$A 483, 509 (2008).
\bibitem{bib:enhance} V. Lefranc {\it et al.}, JCAP09, 043 (2016).
\bibitem{bib:wino} G.F. Giudice, {\it et al.}, JHEP 12, 027 (1998).
\bibitem{bib:mdm} M. Cirelli, N. Fornengo and A. Strumia, Nucl. Phys. B 753, 178 (2006).


\end{thebibliography}
\end{document}